\documentclass[12pt]{amsart}

\newcommand{\HH}{\mathcal H}
\newcommand{\R}{\mathbb R}

\begin{document}

\title{Excitations propagating along surfaces}
\author{A.~V.~Stoyanovsky}
\date{}
\address{Moscow Center for Continuous Mathematical Education}
\email{stoyan@mccme.ru}
\maketitle

This paper is a development of the note [1].
The purpose of the paper is to deduce a number of equations which describe excitations
propagating along $n$-dimensional surfaces in $\R^N$. Usual excitations in wave theory
propagate along one-dimensional trajectories. In the present paper the role of the medium
of propagation of excitations is played by the infinite
dimensional space of $(n-1)$-dimensional surfaces in $\R^N$. The role of rays
is played by $n$-dimensional solution surfaces of a variational problem.
Such a generalization of wave theory can be useful in quantum field theory [2].

The paper consists of three sections. In \S1 the formula for variation
of action given by an $n$-dimensional integral is recalled. In \S2 analogs of
the Hamilton--Jacobi equation and of the canonical Hamilton equations are deduced.
Besides that, the analogs of the canonical Hamilton equations are identified with
the equations of characteristics for the analog of the Hamilton--Jacobi equation.
The analog of the Hamilton--Jacobi equation for $n$-dimensional variational problems
has been found in particular cases by many authors, see, for example, [3]
and references therein. The book [3] also contains a theory of integration of
the analog of the Hamilton--Jacobi equation, but the equations of characteristics
are written in [3] in a form somewhat different from that of the present paper.
In \S3 an analog of the wave Schrodinger equation is deduced, and an analog
of the transport equation is written out in the first order of
quasiclassical approximation. Besides that, the Cauchy problem for the analog of the
Hamilton--Jacobi equation is formulated and solved.

\section{The formula for variation of action}

\subsection{The main formula}
Consider the variational problem for the integral
\begin{equation}
J=\int_D F(x^1,\ldots,x^n,z^1,\ldots,z^m,z^1_{x^1},\ldots,z^m_{x^n})\,
dx^1\ldots dx^n.
\end{equation}
Here $x^1,\ldots,x^n$ are independent variables, $z^1,\ldots,z^m$ are dependent
variables, $z^i_{x^j}=\frac{\partial z^i}{\partial x^j}$,
and integration goes over an $n$-dimensional surface $D$ with
boundary $\partial D$ in $\R^{m+n}$.

The Euler--Lagrange equations for the functions $z^i(x)$ read
\begin{equation}
F_{z^i}-\sum_j\frac{\partial}{\partial x^j}F_{z^i_{x^j}}=0,\ \ \ i=1,\ldots,m.
\end{equation}
Here $F_{z^i}$ and $F_{z^i_{x^j}}$ denote the partial derivatives of the function $F$
with respect to the corresponding arguments.

Assume that for each $(n-1)$-dimensional parameterized surface $C$
in $\R^{m+n}$ given by the equations
\begin{equation}
x^j=x^j(s^1,\ldots,s^{n-1}),\ \ z^i=z^i(s^1,\ldots,s^{n-1})
\end{equation}
and sufficiently close to a fixed $(n-1)$-dimensional surface,
there exists a unique $n$-dimensional surface $D$ with boundary $\partial D=C$
which is
an extremal of the integral (1), i.e., the graph of a solution
to the Euler--Lagrange equations (2). Denote by $S=S(C)$ the value of the integral (1)
over the surface $D$.
The goal of this section is to deduce the well known formula for variation
of the functional $S$.

To this end, let us represent the integral (1) in the parametric form with the
parameters $s^1,\ldots,s^{n-1},t$, assuming that the boundary $\partial D$
corresponds to one or several fixed values of the variable $t$.
We obtain
\begin{equation}
\begin{aligned}{}
J=\int_D &F(x^1(s,t),\ldots,x^n(s,t),z^1(s,t),\ldots,z^m(s,t),\\
&\frac{\partial(z^1,x^2,\ldots,x^n)}{\partial(t,s^1,\ldots,s^{n-1})}:
\frac{\partial(x^1,\ldots,x^n)}{\partial(t,s^1,\ldots,s^{n-1})},
\ldots,\\
&\frac{\partial(x^1,\ldots,x^{n-1},z^m)}{\partial(t,s^1,\ldots,s^{n-1})}:
\frac{\partial(x^1,\ldots,x^n)}{\partial(t,s^1,\ldots,s^{n-1})})\\
&\frac{\partial(x^1,\ldots,x^n)}{\partial(t,s^1,\ldots,s^{n-1})}\,dtds,
\end{aligned}
\end{equation}
where
$\frac{\partial(x^1,\ldots,x^n)}{\partial(s^1,\ldots,s^n)}=\left|
\frac{\partial x^j}{\partial s^i}\right|$ is the Jacobian,
$ds=ds^1\ldots ds^{n-1}$. Denote the integrand in the latter
integral by $\Phi(x^j,z^i,x^j_{s^k},z^i_{s^k},x^j_t,z^i_t)$. Then we have
(omitting the summation sign over repeating indices)
\begin{equation}
\begin{aligned}{}
&\delta J=\int_D(\Phi_{x^j}\delta x^j+\Phi_{z^i}\delta z^i+\Phi_{x^j_{s^k}}
\delta x^j_{s^k}+\Phi_{x^j_t}\delta x^j_t+\Phi_{z^i_{s_k}}\delta z^i_{s^k}
+\Phi_{z^i_t}\delta z^i_t)\,dtds\\
&=\int_D\left[\left(\Phi_{x^j}-\frac{\partial}{\partial s^k}\Phi_{x^j_{s^k}}
-\frac{\partial}{\partial t}\Phi_{x^j_t}\right)\delta x^j+
\left(\Phi_{z^i}-\frac{\partial}{\partial s^k}\Phi_{z^i_{s^k}}
-\frac{\partial}{\partial t}\Phi_{z^i_t}\right)\delta z^i\right]dtds\\
&+\int_D\left[\frac{\partial}{\partial s^k}\left(\Phi_{x^j_{s^k}}
\delta x^j+\Phi_{z^i_{s_k}}\delta z^i\right)+
\frac{\partial}{\partial t}\left(\Phi_{x^j_t}
\delta x^j+\Phi_{z^i_t}\delta z^i\right)\right]dtds.
\end{aligned}
\end{equation}
It is not difficult to see that the first integral here vanishes due to the
Euler--Lagrange equations (2). And the second integral equals,
by the Stokes theorem,
\begin{equation}
\delta S=\int_{\partial D}\left(\Phi_{x^j_t}
\delta x^j+\Phi_{z^i_t}\delta z^i\right)\,ds.
\end{equation}
Hence
\begin{equation}
\begin{aligned}
\frac{\delta S}{\delta z^i(s)}&=p^i(s),\\
\frac{\delta S}{\delta x^j(s)}&=-H^j(s),
\end{aligned}
\end{equation}
where
\begin{equation}
\begin{aligned}
p^i=\Phi_{z^i_t}&=\sum_l(-1)^{l+1}F_{z^i_{x^l}}
\frac{\partial(x^1,\ldots,\widehat{x^l},\ldots,x^n)}{\partial(s^1,\ldots,s^{n-1})},\\
H^j=-\Phi_{x^j_t}&=\sum_{l\ne j}(-1)^{l+1}F_{z^i_{x^l}}z^i_{x^j}
\frac{\partial(x^1,\ldots,\widehat{x^l},\ldots,x^n)}{\partial(s^1,\ldots,s^{n-1})}\\
&+(-1)^{j+1}(F_{z^i_{x^j}}z^i_{x^j}-F)
\frac{\partial(x^1,\ldots,\widehat{x^j},\ldots,x^n)}{\partial(s^1,\ldots,s^{n-1})}.
\end{aligned}
\end{equation}
Here the cap over a variable means that the variable is omitted; the summation sign
over the index $i$ repeated twice is omitted.
Note that in the formula for $H^j$, the coefficients before the Jacobians
coincide up to sign with the components of the energy-momentum tensor.
Also note that the quantities $p^i$ and $H^j$ depend on the numbers $x^j_t$, $z^i_t$
only through the numbers $z^i_{x^j}$ characterizing the tangent space
to the $n$-dimensional surface $D$.

Thus, we obtain the formula
\begin{equation}
\delta S=\int_C \left(\sum p^i\delta z^i-\sum H^j\delta x^j\right)\,ds
\end{equation}
analogous to the formula for the differential of the action function in
classical mechanics. This formula easily implies the Noether theorem, as in
classical mechanics.

\subsection{The properties of the quantities $p^i$ and $H^j$}

It is easy to see that the quantities $p^i$ and $H^j$ satisfy the relations
\begin{equation}
p^iz^i_{s^k}-H^jx^j_{s^k}=0,\ \ \ k=1,\ldots,n-1,
\end{equation}
and
\begin{equation}
p^iz^i_t-H^jx^j_t=\Phi.
\end{equation}
The latter equation is just the Euler theorem on homogeneous functions, since
the function $\Phi$ is homogeneous of degree 1 in the variables
$z^i_t$, $x^j_t$.

Let us call by the tangent element the data consisting of an $(n-1)$-dimensional
surface $C$ and a family of $n$-dimensional planes tangent to the surface
at each point of the surface. These planes can be characterized by $mn$
variables $z^i_{x^j}$ satisfying $m(n-1)$ equations
\begin{equation}
z^i_{x^j}x^j_{s^k}=z^i_{s^k},\ \ \ k=1,\ldots,n-1.
\end{equation}

On the other hand, these planes or, more precisely, the dual conormal lines
can be characterized by $m+n$
variables $p^i$, $H^j$ subject to $(n-1)$ equations (10).
One passes from the former set of variables to the latter one
by the duality transform (8).

A count of parameters shows that the quantities $p^i$, $H^j$ should satisfy
one more equation. To write it explicitly, assume for a moment that the function
$\Phi$ of the variables $x^j_t,z^i_t$ is convex, so that it yields a norm on
the quotient space of the space of vectors $(z^i_t,x^j_t)$ by the transformations
\begin{equation}
\begin{aligned}
x^j_t&\to x^j_t+\sum a_k x^j_{s^k},\\
z^i_t&\to z^i_t+\sum a_k z^i_{s^k}
\end{aligned}
\end{equation}
with arbitrary $a_k$.
Denote by $\HH$ the norm on the dual space of vectors $(p^i,H^j)$
satisfying relations (10), i.e., for arbitrary $(p^i,H^j)$ satisfying (10) put
\begin{equation}
\HH(p^i,-H^j)=\max\limits_{\Phi(z^i_t,x^j_t)=1}p^iz^i_t-H^jx^j_t.
\end{equation}
Then it is not difficult to deduce the following relation between the quantities
$p^i$, $H^j$ (8):
\begin{equation}
\HH(p^i,-H^j)-1=0.
\end{equation}
The function $\HH$ is determined as a function of $m+n$ variables not uniquely,
but only up to adding a product of any function with the left-hand side
of one of equations (10).
In the general case of not necessarily convex function $\Phi$,
let us mean by equation (15) the $n$-th equation for
the quantities $p^i,H^j$.

Using the $n$ equations (10) and (15), one can (in general) express the quantities $H^j$
as functions of $p^i$ (and of $x^l$, $z^i$, $x^l_{s_k}$, $z^i_{s_k}$):
\begin{equation}
H^j=H^j(x^l,z^i,x^l_{s^k},z^i_{s^k},p^i),\ \ \ j=1,\ldots,n.
\end{equation}

The change of variables (8), from $mn$ variables $z^i_{x^j}$ subject to $m(n-1)$ equations
(12) and from the function $F$ to $m$ variables $p^i$ and to $n$ functions $H^j$ (16),
can be called the modified Legendre transform.
Up to notations it coincides with the usual Legendre transform. Indeed,
if instead of the variables $z^i_{x^j}$ we consider
$m$ variables $z^i_t$, then the function $H^jx^j_t$ of the variables $p^i$ will be the
(usual) Legendre transform of the function $\Phi$, see formulas (8) and (11). Therefore,
since the Legendre transform is a duality, we obtain the formulas
\begin{equation}
z^i_t=z^i_{x^j}x^j_t=(H^jx^j_t)_{p^i}=H^j_{p^i}x^j_t.
\end{equation}
Besides that, if we differentiate equations (10) with respect to the variable $p^i$,
then we obtain the system of equations
\begin{equation}
z^i_{s^k}=z^i_{x^j}x^j_{s^k}=H^j_{p^i}x^j_{s^k},\ \ \ k=1,\ldots,n-1
\end{equation}
which, together with (17), gives
\begin{equation}
z^i_{x^j}=H^j_{p^i}.
\end{equation}

\subsection{Geometric optics in the space of $(n-1)$-dimensional surfaces}

The above formulas become more clear if one has in mind the following geometric
picture. Consider the space whose points $C$ are $(n-1)$-dimensional parameterized
surfaces (3) in $\R^{m+n}$. Consider propagation of light in this space by the Fermat
principle, the time of propagation of light along a trajectory $C(t)$,
$t_0\le t\le t_1$, being given by the $n$-dimensional integral (4).
Let us call by the light rays the trajectories $C(t)$ for which integral (4) takes
a stationary value. Here one has the following essential difference with
usual geometric optics: there are not one but infinitely many rays passing through
two given points. These rays correspond to various parameterizations of
$n$-dimensional solution surface $D$ of the variational problem.

Let us fix a point $C_0$, and denote by $S(C)$ the time of propagation of light
from the point $C_0$ to the point $C$. Let us call a level hypersurface $S(C)=T$
of the function $S$ by a wave front emitted by the point $C_0$ at the time
$T$, and denote this front by $W(T)$. Then it is natural to assume that
the following Huyghens principle holds: the wave front $W(T+\Delta T)$ is the
common tangent hypersurface to the fronts emitted by the points of the
wave front $W(T)$ at the time
$\Delta T$ (which may be called spheres of radius $\Delta T$).
Here again one meets the following difference with usual geometric optics.
At a given point of the front $W(T+\Delta T)$, there are infinitely many
spheres tangent to the front. These spheres correspond to partitions of the surface
$D$ of ``square'' $T+\Delta T$ into two pieces of ``squares''
$T$ and $\Delta T$. Similarly, each sphere is tangent to the front $W(T+\Delta T)$
at infinitely many points.

Now if we take $\Delta T$ infinitely small, then we obtain the following picture.
Consider a sphere of an infinitely small radius $dT$ with the center at a point
$C$ of the front $W(T)$. Then the function $S$ achieves its maximum at the intersection
points of the sphere with the rays passing through the points
$C_0$ and $C$. Hence the tangent hyperplane to the wave front
at the point $C$ is parallel to the tangent hyperplane to the sphere at any of these
intersection points. In particular, the slope of the tangent hyperplane
does not depend on a concrete point of the sphere, and depends only on the tangent
planes to the surface $D$ (the tangent element, see Subsection 1.2). From these
considerations one can deduce formulas (7,8).

\section{Analogs of Hamilton's formulas}

\subsection{Analog of the Hamilton--Jacobi equation}
Substituting (7) into equations (10,15) or into (16), we obtain
\begin{equation}
\begin{aligned}{}
\frac{\delta S}{\delta z^i(s)}z^i_{s^k}+\frac{\delta S}{\delta x^j(s)}
x^j_{s^k}&=0,\ \ \ k=1,\ldots,n-1,\\
\HH\left(\frac{\delta S}{\delta z^i(s)},\frac{\delta S}{\delta x^j(s)}
\right)-1&=0,
\end{aligned}
\end{equation}
or
\begin{equation}
\frac{\delta S}{\delta x^j(s)}+H^j\left(x^l,z^i,x^l_{s^k},z^i_{s^k},
\frac{\delta S}{\delta z^i(s)}\right)=0,\ \ \ j=1,\ldots,n.
\end{equation}
The system of equations (20) or (21), relating the variational derivatives
of the functional $S$ at the same point
$s$, can be naturally called the analog of the Hamilton--Jacobi equation.
The first $n-1$ equations of the system (20) correspond to the fact that the
function $S$ does not
depend on the parameterization of the surface $C$.

\medskip
{\bf Example} (the scalar field in two-dimensional space-time) [1].
Let $F(x,y,z,z_x,z_y)=\frac{1}{2}(z_x^2-z_y^2)+p(z)$, where
$p(z)=\frac{m^2}{2}z^2+\ldots$ is a polynomial in $z$.
A calculation gives the following analog of the Hamilton--Jacobi equation:
\begin{equation}
\begin{aligned}{}
&x_s\frac{\delta S}{\delta x}+y_s\frac{\delta S}{\delta y}+
z_s\frac{\delta S}{\delta z}=0,\\
\frac{1}{2}\left(\left(\frac{\delta S}{\delta z}\right)^2+
z_s^2\right)&+\left(x_s^2-y_s^2\right)p(z)+x_s\frac{\delta
S}{\delta y}+y_s\frac{\delta S}{\delta x}=0.
\end{aligned}
\end{equation}

\subsection{Analogs of the canonical Hamilton equations}

Let us assume again that the surface $D$ is parameterized by the coordinates
$s_1,\ldots,s_{n-1},t$, and let us find the equations expressing the dependence
of the variables $p^i,z^i$ on $t$, assuming that the dependence of $x^j$ on $(s,t)$
is given.

To this end, consider the integral
\begin{equation}
\int\!\int (p^iz^i_t-H^jx^j_t)\,dsdt
\end{equation}
as a functional of functions $p^i,z^i$, in which $H^j$ is the function (16),
and let us write the Euler--Lagrange equations for this functional.
Varying $p^i$, we obtain equation (17). It implies that
the function $p^iz^i_t-H^jx^j_t$ is the Legendre transform of the function
$H^jx^j_t$ with respect to the variables $p^i$. Hence it coincides with
the function $\Phi$ defined in \S1. Hence the Euler--Lagrange equations
for the integral (23) are equivalent to the Euler--Lagrange equations for
the function $\Phi$ and therefore to the Euler--Lagrange equations (2).
It remains to vary the integral (23) with respect to
$z^i$, which yields the following form of the field equations:
\begin{equation}
\begin{aligned}{}
z^i_t&=\frac{\delta}{\delta p^i(s)}\int H^jx^j_t(s)\,ds, \\
p^i_t&=-\frac{\delta}{\delta z^i(s)}\int H^jx^j_t(s)\,ds.
\end{aligned}
\end{equation}
These equations can be naturally called analogs of the canonical Hamilton
equations.

\subsection{Analogs of the canonical Hamilton equations as the equations
of characteristics for the analog of the Hamilton--Jacobi equation}
The analog of the Hamilton--Jacobi equation can be written in the form (9),
where $H^j$ are the functions (16). In this form, the analog of the
Hamilton--Jacobi equation is a system of equations for the coordinates
$p^i(s),H^j(s)$ of the tangent hyperplane to the graph of a solution
$S$.
At a fixed point $(x^j(s),z^i(s),S)$ of the graph, this system gives
a family of tangent hyperplanes parameterized by functions $p^i(s)$.
The common tangent hypersurface to this family is the analog of the Monge
cone (see [4], Ch.~II). It is given by the equations obtained by
taking the variational derivative of equation (9) with respect to $p^i(s)$,
i.e.,
\begin{equation}
z^i_t(s)=H^j_{p^i}x^j_t(s),
\end{equation}
where $z^i_t=\frac{\delta z^i}{\delta t}$, $x^j_t=\frac{\delta x^j}{\delta t}$.
These are equations (17) forming the first half of the analogs of the canonical
Hamilton equations.

Let us call a 4-tuple of functions $(x^j(s),z^i(s),p^i(s),H^j(s))$, where
$H^j(s)=-\frac{\delta S}{\delta x^j(s)}$,
$p^i(s)=\frac{\delta S}{\delta z^i(s)}$,
by an integral element of a solution to the analog of the Hamilton--Jacobi equation.
An integral element gives a point $C$ and the coordinates of the tangent
hyperplane to the graph of the solution at this point.
Assume we are given a one-parameter family of integral elements
$(x^j(s,t),z^i(s,t),p^i(s,t),H^j(s,t))$ of a given solution $S$. Let us call
the family
focal (as in the usual theory of the Hamilton--Jacobi equation)
if the analog (25) of the Monge equations holds. Then one must also have the
equations
\begin{equation}
\begin{aligned}{}
p^i_t&=\frac{\partial}{\partial t}\frac{\delta S}{\delta z^i(s)}
=\int x^j_t(s')\frac{\delta^2 S}{\delta z^i(s)\delta x^j(s')}\,ds'
+\int z^{i'}_t(s')\frac{\delta^2 S}{\delta z^i(s)\delta z^{i'}(s')}\,ds'\\
&=-\int x^j_t(s')\frac{\delta}{\delta z_i(s)}H^j(s',p^{i'}(z,x))\,ds'
+\int H^j_{p^{i'}}x^j_t(s')\frac{\delta}{\delta z_i(s)}p^{i'}(s')\,ds'\\
&=-\frac{\delta}{\delta z^i(s)}\int H^jx^j_t(s')\,ds',
\end{aligned}
\end{equation}
where in the last integral the functions $H^j$ are considered as functions
of independent variables
$z^i,p^i$.
The latter computation is somewhat formal but it is not difficult to
make it rigorous.

Thus, a focal family of integral elements belongs to the graph of a solution
$S$ only if equations (25,26), i.e., the analogs of the canonical Hamilton
equations, hold. If we pass from integral elements $(x^j,z^i,p^i,H^j)$
to tangent elements $(x^j,z^i,z^i_{x^j})$, then the last condition means that
this family of tangent elements covers a surface $z=z(x)$ which is the graph of
a solution to the Euler--Lagrange equations (2). So the analogs of the canonical
Hamilton equations can be naturally called the equations of characteristics
for the analog of the Hamilton--Jacobi equation. And solution surfaces of the
Euler--Lagrange equations (2) can be called analogs of rays.

\section{Analog of the Schrodinger equation}

\subsection{Analog of the Schrodinger equation}

Suppose that the functions $H^j$ (16) are polynomials in the variables $p^i$.
Let us make the following substitution in the analog (21) of the Hamilton--Jacobi
equation:
\begin{equation}
\begin{aligned}{}
&\frac{\delta S}{\delta x^j(s)} \to -ih\frac{\delta}{\delta x^j(s)},\\
&\frac{\delta S}{\delta z^{i'}(s)}=p^{i'} \to -ih\frac{\delta}{\delta z^{i'}(s)}.
\end{aligned}
\end{equation}
Here $i$ is the imaginary unit, $h$ is a very small constant.
We obtain a system of linear variational differential equations
which can be naturally called the analog of the Schrodinger equation:
\begin{equation}
-ih\frac{\delta \Psi}{\delta x^j(s)}+H^j\left(x^l,z^{i'},x^l_{s^k},z^{i'}_{s^k},
-ih\frac{\delta}{\delta z^{i'}(s)}\right)\Psi=0,\ \ \ j=1,\ldots,n.
\end{equation}
Here $\Psi=\Psi(C)$ is the unknown complex-valued functional of functions
$x^j(s),z^{i'}(s)$; it is assumed that in the polynomials $H^j$ the operators
$-ih\frac{\delta}{\delta z^{i'}(s)}$ stand to the right of the coefficients of
the polynomial.

The system (28) can be also written in the form of type (20) if we assume that
the left-hand side of equation (15) is a polynomial with respect to the variables
$H^j,p^{i'}$:
\begin{equation}
\begin{aligned}{}
\frac{\delta \Psi}{\delta z^{i'}(s)}z^{i'}_{s^k}+\frac{\delta \Psi}{\delta x^j(s)}
x^j_{s^k}&=0,\ \ \ k=1,\ldots,n-1,\\
\left[\HH\left(-ih\frac{\delta}{\delta z^{i'}(s)},-ih\frac{\delta}{\delta x^j(s)}
\right)-1\right]\Psi&=0.
\end{aligned}
\end{equation}
The first $n-1$ equations of the system (29) can be replaced by the condition that
the functional $\Psi(C)$ does not depend on the parameterization of a surface $C$.

\medskip
{\bf Example.} In the example considered at the end of Subsection 2.1, we obtain
the following analog of the Schrodinger equation:
\begin{equation}
\begin{aligned}{}
\frac{1}{2}\left(-h^2\frac{\delta^2\Psi}{\delta z(s)^2}+
z_s^2\Psi\right)&+\left(x_s^2-y_s^2\right)p(z)\Psi
-ih\left(x_s\frac{\delta\Psi}{\delta y(s)}+y_s\frac{\delta\Psi}{\delta
x(s)}\right)=0;\\
x_s&\frac{\delta\Psi}{\delta x(s)}+
y_s\frac{\delta\Psi}{\delta y(s)}+z_s\frac{\delta\Psi}{\delta
z(s)}=0.
\end{aligned}
\end{equation}

\subsection{The quasiclassical approximation}
Let us substitute into the system of equations (28) a functional of the form
\begin{equation}
\Psi(C)=a(C)e^{\frac{iS(C)}{h}}.
\end{equation}
Equating to zero the principal part with respect to $h$, we obtain the analog (21)
of the Hamilton--Jacobi equation for the functional $S$. Equating to zero the coefficient
before $h$, we obtain the analog of the transport equation:
\begin{equation}
\frac{\delta a}{\delta x^j(s)}+H^j_{p_i}\left(x^l,z^{i'},x^l_{s^k},
z^{i'}_{s^k},p^{i'}\right)\frac{\delta a}{\delta z^i(s)}=0,\ \ \ j=1,\ldots,n,
\end{equation}
where $p^i=\frac{\delta S}{\delta z^i(s)}$.
This equation can be interpreted in the following way.
Consider a one-parameter family of integral elements of the solution $S$
of the analog of the Hamilton--Jacobi equation
satisfying the equations of characteristics, i.e.,
the analogs of the canonical Hamilton equations.
Then, as shown above, the corresponding tangent elements cover an $n$-dimensional
surface $D$: $z=z(x)$ which is a solution to the Euler--Lagrange equations.
Also relation (19) holds. Therefore, the analog of the transport equation
means that the ``amplitude'' $a(C)$ is constant along the one-parameter family.

Suppose that the solution to the Cauchy problem for the Euler--Lagrange equations
(2) with the given tangent element for $t=t_0$ is unique. Then for any other
one-parameter family passing through the same surface $C_0=C(t_0)$,
the corresponding tangent elements cover the same surface $D$.
Suppose that the analog of the Monge equations (25) has a solution $z^i(s,t)$
for any functions $x^j(s,t)$ and the initial conditions $(x^j(s,t_0),z^i(s,t_0))$.
Then any $(n-1)$-dimensional surface $C$ on the surface $D$
can be included into a one-parameter family.
Hence, the integral element at the point $C$ corresponds to the tangent element
tangent to the surface $D$, and the value of the function $S$ at the point
$C$ is given by the formula
\begin{equation}
S(C)=S(C_0)+\int_{D_1}\Phi\,dsdt,
\end{equation}
where $D_1$ is the piece of the surface $D$ bounded by the $(n-1)$-dimensional
surfaces $C_0$ and $C$. Formula (33) follows from formulas (9),(11) and from the argument
of Subsect. 2.2.
We obtain that the function $a(C)$ does not depend on an $(n-1)$-dimensional surface
$C$ on the surface $D$, but depends only on the surface $D$ itself.
This is the analog of the statement from quasiclassical approximation of
quantum mechanics or from short-wave limit of wave optics that the amplitude of the
wave function is constant along the rays.

In [3] the sets of integral elements corresponding to all tangent elements
tangent to the same solution surface $D$ of the Euler--Lagrange equations
are called the characteristics of the analog of the Hamilton--Jacobi equation.
Thus, we obtain that in general for a solution $S$
the space of surfaces $C$ is covered by a family of pairwise non-intersecting
projections of characteristics along which the function $a(C)$ is constant.
The arising family of surfaces $D$ can be called the analog of the notion of
field of extremals in one-dimensional variational calculus.

\subsection{The Cauchy problem for the analog of the Hamilton--Jacobi equation}
Let us pose the problem to find a solution $S$ of equations (21) which takes
the given values $S(C_0)=U(C_0)$ on the surfaces $C_0$ of the form
\begin{equation}
x^j(s)=b^j(s),\ \  z^i(s)=a^i(s),
\end{equation}
where $b^j(s)$ are the given functions, and $a^i(s)$ are arbitrary functions.
To construct solution of this Cauchy problem, note that at each point
$C_0$ one can define an integral element
$p^i(s)=\frac{\delta U}{\delta a^i(s)}$, and therefore a tangent element.
For a given point $C$ let us pass through this point a projection of a characteristic,
i.~e., let us find $C_0$ such that the solution surface $D$ of the Euler--Lagrange
equations passing through the surfaces $C_0$ and $C$, would be tangent to the
above constructed tangent element along the surface $C_0$. Assume that such
$C_0$ and $D$ exist and are unique. After that let us apply formula (33).

The thus constructed function $S$ can be obtained also using the construction of
common tangent hypersurface generalizing the Huyghens principle (see 1.3).
Assume that for each surface $C$ (3) sufficiently close to a fixed $(n-1)$-dimensional
surface, and for each surface $C_0$ (34) sufficiently close to some other
$(n-1)$-dimensional surface, there exists a unique $n$-dimensional solution surface
$D$ of the Euler--Lagrange equations with the boundary $C_0\cup C$.
Denote by $I(C_0,C)$ the integral of the function $\Phi$ over the surface $D$.
Then for each $C_0$ the function $S_{C_0}(C)=I(C_0,C)+U(C_0)$ is a solution of the
analog of the Hamilton--Jacobi equation (full integral, cf. [1]). The graph of
the function $S(C)$ is the common tangent hypersurface to the family of the graphs
of the functions $S_{C_0}(C)$ for all $C_0$. Indeed, the common tangent
hypersurface is obtained by expressing the quantities $a_i(s)$ as functions of
$x^j(s'),z^i(s')$ from the equations
\begin{equation}
\begin{aligned}{}
&S(C)=I(C_0,C)+U(C_0),\\
\frac{\delta I}{\delta a^i(s)}&+\frac{\delta U}{\delta a^i(s)}=0,
\ \ \ i=1,\ldots,m.
\end{aligned}
\end{equation}
These equations imply that the function $S(C)$ coincides with the one constructed above.
Besides that, its variation at the point $C$ coincides with the variation of the
function $S_{C_0}(C)$. Therefore, the function $S$ satisfies the analog of the
Hamilton--Jacobi equation, and its integral element at the point $C$
corresponds to the tangent element tangent to the surface $D$.

The analog of the field of extremals for the solution $S$ is the family of
solution surfaces $D$ of the Cauchy problem for the Euler--Lagrange equations with the
given tangent elements along the surfaces $C_0$. The characteristics are
the varieties along which the graph of the common tangent hypersurface touches
the graphs of solutions $S_{C_0}(C)$ from the full integral.

Thus, we see that the theory of integration of the Hamilton--Jacobi equation
from [4], Ch.~2 can be generalized to multidimensional variational problems.

\end{document}